# A planetary system as the origin of structure in Fomalhaut's dust belt

Paul Kalas[1], James R. Graham[1], Mark Clampin[2]

[1]*Astronomy Department, University of California, Berkeley, CA 94720 USA*

[2]*Goddard Space Flight Center, Greenbelt, MD 20771 USA*

The Sun and >15 percent of nearby stars are surrounded by dusty debris disks that must be collisionally replenished by asteroids and comets, as the dust would otherwise be depleted on <$10^7$ yr timescales (ref. 1). Theoretical studies show that disk structure can be modified by the gravitational influence of planets[2-4], but the observational evidence is incomplete, at least in part because maps of the thermal infrared emission from disks have low linear resolution (35 AU in the best case[5]). Optical images provide higher resolution, but the closest examples (AU Mic and Beta Pic) are edge-on[6,7], preventing the direct measurement of azimuthal and radial disk structure that is required for fitting theoretical models of planetary perturbations. Here we report the detection of optical light reflected from the dust grains orbiting Fomalhaut (HD 216956). The system is inclined 24° away from edge-on, enabling the measurement of disk structure around its entire circumference, at a linear resolution of 0.5 AU. The dust is distributed in a belt 25 AU wide, with a very sharp inner edge at a radial distance of 133 AU, and we measure an offset of 15 AU between the belt's geometric centre and Fomalhaut. Taken together, the sharp inner edge and offset demonstrate the presence of planet-mass objects orbiting Fomalhaut.

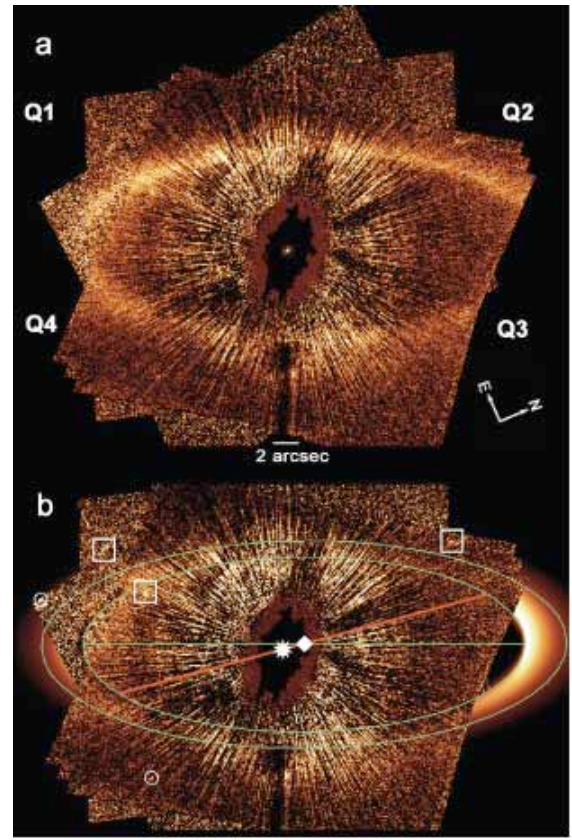

**Figure 1:** Optical detection of dust in Fomalhaut's Kuiper Belt. (a) False-color coronagraphic image (Methods). North is 66° clockwise from vertical. (b) Residual image after subtracting a model of a dust belt (Methods). Green lines trace the projected boundaries of the detected belt (133 - 158 AU). The white diamond and asterisk mark the centres of the belt and star, respectively. The horizontal green line traces the belt semi-major axis, whereas the red line traces the vector between the belt and star centres. White boxes and circles mark extended objects and background stars, respectively.

Fomalhaut's circumstellar dust was discovered at thermal infrared wavelengths with the Infrared Astronomical Satellite[8,9], and maps of dust emission at sub-millimetre wavelengths with 7.5" – 14" angular resolution suggested a ring of material residing between 100 and 140 AU radius[10,11]. Fomalhaut is only 7.7 parsecs (1 pc = 3.3 light years) from the Sun. Using the Advanced Camera for Surveys (ACS) aboard the Hubble Space Telescope (HST), we detect for the first time Fomalhaut's dust complex at optical wavelengths, with angular resolution 100 times sharper than previous sub-mm maps (Fig. 1; Methods). Fomalhaut is surrounded by a narrow dust belt inclined to our line of sight and with significant asymmetry in its azimuthal brightness distribution. We perform a non-linear least-squares fit to the ellipse traced by the brightest points along the belt. The semi-major and semi-minor axes are 140.7 ± 1.8 AU and 57.5 ± 0.7 AU, respectively. The position angle of the ellipse is PA = 156.0° ± 0.3°, with inclination to the line of sight $i$ = 65.9° ± 0.4° (assuming the structure is intrinsically circular). These values are consistent with the belt properties previously inferred via modeling of sub-mm data[11].

The elliptical fit to Fomalhaut's belt has a projected offset from the star 13.4 ± 1.0 AU at PA = 350.4°. Assuming that the star and belt are coplanar, the projected offset translates to 15.3 AU in the plane of the belt. A Keplerian orbit with non-zero eccentricity traces an ellipse where the star is at one focus and the centre of the ellipse is offset by a value, $f$, equivalent to the semi-major axis, $a$, times the eccentricity, $e$. If we assume Fomalhaut's belt is non-circular, then the observed offset and semi-major axis give $e$ = 0.11± 0.01. Periastron is at PA = 170°. The non-circularity

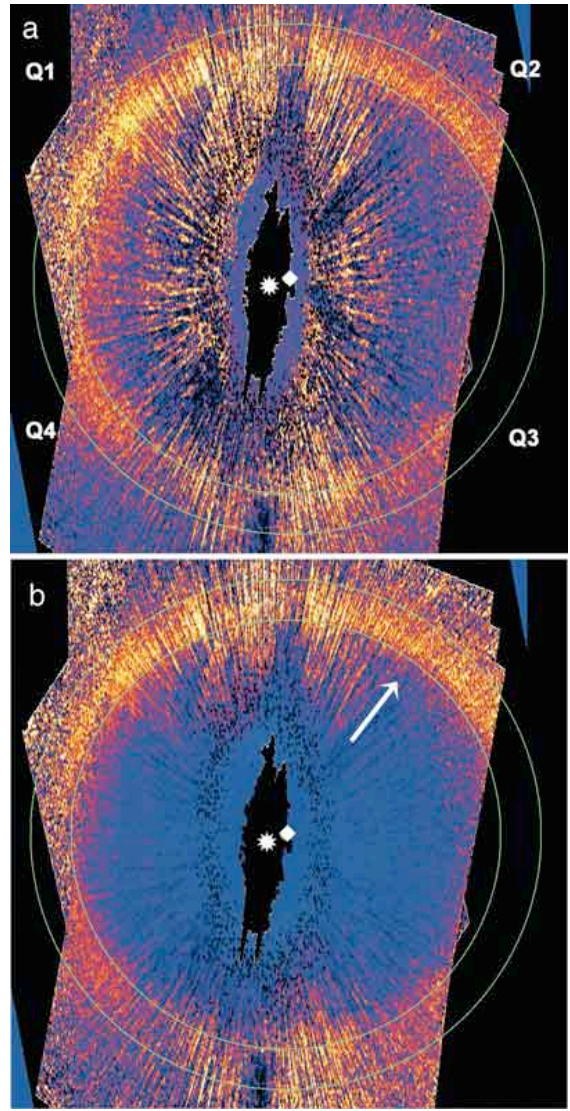

**Figure 2** False color representations of the de-projected belt image, with symbols as in Fig. 1. (**a**) The observed, de-projected surface brightness. (**b**) The surface brightness scaled by distance squared from Fomalhaut, correcting for the inverse square dilution of radiation. The stellar flux normalization is chosen so that it is unity at the center of the ring (140 AU). If the Fomalhaut dust belt consists of grains of uniform optical scattering properties, then (**b**) maps the material surface density for an optically thin distribution. A white arrow indicates the direction of the radial cut in Fig. 3.

of the structure leads to a small revision for the inclination, $i = 65.6° \pm 0.4°$ (ref. 12).

The offset of the star relative to the dust belt and an asymmetric scattering phase function produce the azimuthal brightness asymmetries. In the rotated frame shown in Fig. 1 we refer to the upper left quadrant of the belt as quadrant one (Q1), with Q2, Q3 and Q4 clockwise from Q1. We sample the surface brightness in these quadrants at angles 25° away from the horizontal axis, giving $20.6 \pm 0.1$, $21.0 \pm 0.1$, $21.5 \pm 0.1$, and $21.3 \pm 0.3$ mag arcsec$^{-2}$ for Q1, Q2, Q3, and Q4, respectively. The absolute photometry is uncertain by ~0.4 mag arcsec$^{-2}$ due to errors in determining the global sky background. However, the azimuthal variation in relative surface brightness indicated above is a consistent finding in different data sets (Methods). Because the star is closer to the belt in Q1 and Q4, these quadrants appear brighter than the corresponding regions in Q2 and Q3. The fact that half the belt, Q1+Q2, is brighter than the other half, Q3+Q4, points to an asymmetric scattering phase function. To search for azimuthal asymmetry in the belt that does not arise from the centre of symmetry offset, we deproject the data and multiply each element in the belt by the square of its radius from the star (Fig. 2). The phase function asymmetry is still evident, but with no evidence for brightness asymmetry between Q1 and Q2, or between Q3 and Q4 (Fig 2b).

We produce a scattered light model of an optically thin dust belt where we vary the belt geometry and the asymmetry parameter, $g$, in a Henyey-Greenstein phase function (Methods; ref 13). We iteratively subtract models from the data until Fomalhaut's belt disappears from the data (Fig. 1), arriving at model parameters | $g$| = 0.2 and with belt inner edge at 133 AU. The magnitude of the asymmetry parameter is consistent with the observed phase function of Zodiacal light particles ($g \sim +0.2$; forward scattering), and other debris disks (e.g. HD 141569[14]). Particles in the Zodiacal dust cloud have a median size of ~30 µm[15], which is consistent with the ~100 µm grain sizes inferred for Fomalhaut based on model fits to thermal infrared data[11,16]. Moreover, grains smaller than ~7 µm are ejected from the Fomalhaut system by radiation pressure[17,18]. If the grains surrounding Fomalhaut are in fact comparable to our interplanetary dust grains, then the forward scattering direction (the side pointing out of the sky plane) is the top half of the belt (Q1+Q2).

The integrated light from the model belt is 16.2 mag, which corresponds to a total grain scattering cross section $C = 5.5$ x $10^{25}$ cm$^2 / A$, where $A$ is the grain albedo. Given a steady state collisional size distribution $dN = K\ a^{-3.5}$ d$a$, where K is a constant and $a$ is the grain radius, the total disk mass, $M_D$, is

$$M_D = \frac{4\ C\rho\ [a_{max}^{1/2} - a_{min}^{1/2}]}{3\ [a_{min}^{-1/2} - a_{max}^{-1/2}]} \qquad (1)$$

where $a_{min}$ and $a_{max}$ are the minimum and maximum grain sizes, respectively, and $\rho$ is the grain density, assumed to be uniform throughout the belt. Previous sub-mm observations at 450 – 850 µm are most sensitive to grain sizes 0.1 - 1.5 mm, and the Fomalhaut sub-mm data give $M_D = 1.1$ x $10^{26}$ g (ref. 10). Inserting $a_{min} = 0.1$ mm and $a_{max} = 1.5$ mm into eqn. (1), with $\rho = 2.5$ g cm$^{-3}$ and $C = 5.5$ x $10^{25}$ cm$^2$ ($A = 1$), we find $M_D = 7.4$ x $10^{24}$ g. If $A = 0.1 - 0.05$, then $M_D = (0.7 - 1.5)$ x $10^{26}$ g, in agreement with the sub-mm mass. If the optically detected dust and sub-mm detected dust comprise the same grain population, then we conclude that Fomalhaut's belt is composed of relatively dark grains consistent with Kuiper Belt objects and cometary nuclei (Table 1). The total belt mass from $a_{min} = 7$ μm to $a_{max} = 500$ km is (3.4 – 6.8) x $10^{29}$ g (50 - 100 Earth mass) for $A = 0.1$ – 0.05, respectively.

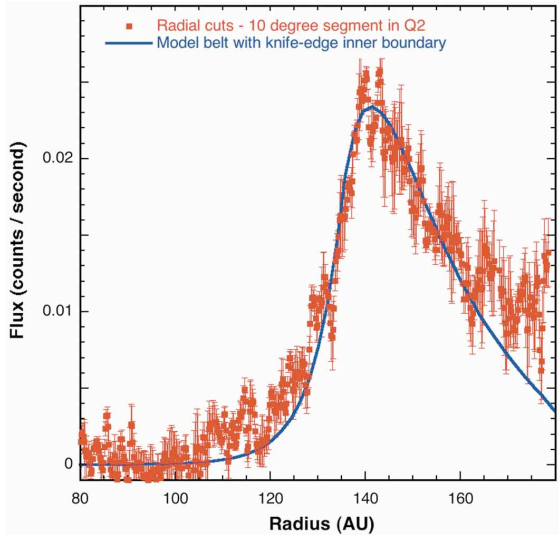

**Figure 3** Radial cut across the belt in the deprojected, illumination-corrected data (Fig. 2). The illumination correction amplifies the sky level and noise at large radii such that the plot shows the sky level beyond ~158 AU. The belt was sampled in 2° increments along a total of 10°. The mean values and the standard deviations per measurement are shown. The blue line is an identical radial cut across the model belt (Methods, Fig. 1). The model belt has sharp inner and outer edges, but the observed edges are softened due to the vertical height of the belt projected onto the sky plane.

The residual image, after model subtraction, emphasizes excess nebulosity interior to the SE lobe (Q1 and Q4) that intrudes inward from the inner edge of the belt to as close as 100 AU radius from the star along the major axis of the belt (Fig. 1b). No comparable excess is detected interior to the NW lobe of the belt. Within equivalent regions between 100 and 120 AU radius along the projected semi-major axis on either side of the belt, the SE interior nebulosity has 1.7 times greater cumulative flux as the NW side (median surface brightness 21.6 ± 0.1 and 22.2 ± 0.1 mag arcsec$^{-2}$ for the SE and NW, respectively). Fomalhaut's inner dust component was recently detected at thermal infrared wavelengths[16,19]. We interpret the inward intrusion of nebulosity in the SE as a tenuous inner dust component that is symmetrical, but detected only in the SE because it is more strongly illuminated due to the star-belt symmetry offset (Fig. 2b).

Fomalhaut's belt appears narrowest in Q2, near apastron. This result contrasts against planetary rings, where the narrowest width occurs at periapse due to rigid precession controlled largely by the self-gravity of ring particles[20]. Future theoretical work will have to consider that self-gravity in Fomalhaut's belt is most likely negligible given the large distance scales and low mass. A radial cut across the belt in Q2 shows a very sharp inner boundary that is consistent with our model belt that assumes a knife-edge inner boundary (Fig. 3). The radial cut shows a plateau at ~140 AU that is 1-2 AU wide. The outer boundary of the belt has a less steep fall-off than the inner edge. However, the model adopted to fit the radial drop-off in surface brightness has a particle number density distribution decreasing as radius$^{-9}$ (Methods). This indicates that the disk outer extent also has a steep drop-off and that the true outer boundary may lie beyond the detection-limited boundary at 158 AU. The model disk adopted here is also very flat, with a 3.5 AU scale height at 133 AU, giving a belt opening angle of ~1.5°.

The sharp inner boundary of the belt and the centre of symmetry offset are known signatures of planet-mass bodies sculpting a debris disk. Fomalhaut's centre of symmetry offset may originate from an eccentric companion that forces the orbital elements of minor bodies into pericentre alignment[21,22]. The net result is a torus of planetesimals with one side uniformly closer to the star. A ~2 AU centre of symmetry offset is also inferred for the dust belt surrounding the star HR 4796A, based on an observed excess of warm grains on one side of the star[21]. A key question is if the sharp inner edge to Fomalhaut's belt is due to planetary resonances, or ejection. Models of our Kuiper Belt dust distribution indicate that the parent bodies accumulate in mean motion resonances with Neptune, producing ring-like features[23]. These models also show that at ~10 AU radius, particles of all sizes are ejected by dynamical interactions with Jupiter and Saturn. We expect that future observations will constrain these possibilities by directly detecting planets around Fomalhaut, or by finding azimuthal dust features that will further elucidate the dynamics of the system.

The emerging picture of Fomalhaut is that of a dust belt with prominent features attributable to a planetary system. Other possible perturbers such as a low mass companion outside of the belt may also modify the structure of the belt. The ~200 Myr age of Fomalhaut[24], is one order of magnitude older than other optically detected debris disk systems such as β Pic[6] and AU Mic[7], and an order of magnitude younger than our Kuiper Belt (Table 1). Further understanding of Fomalhaut's belt may elucidate the epoch of evolution that for our Solar System corresponds to the radial migration of planets due to the gravitational scattering of minor bodies, the recent formation of Pluto-sized objects at the outer edge of the system, the inward delivery of primordial icy material to the surfaces and atmospheres of terrestrial planets, and the formation of the Kuiper Belt and Oort cloud.

**Acknowledgements:** This research is based on observations with the NASA/ESA Hubble Space Telescope, which is operated by the Association of Universities for Research in Astronomy. P.K. acknowledges support from the Space Telescope Science Institute (STScI) and NASA's Origins of Solar Systems program. P.K. and J.R.G. also thank the NSF Center for Adaptive Optics, managed by the University of California, Santa Cruz. The authors also thank STScI for observations allocated under the Director's Discretionary Time program.

**Methods**

Because Fomalhaut is one of the brightest stars in the sky ($m_V = 1.16$ mag), we utilized the ACS coronagraphic camera with a 1.8" diameter occulting spot to artificially eclipse the star. A red broadband filter (F814W, $\lambda_c = 833$ nm, $\Delta\lambda = 251$ nm), comparable to the Johnson-Cousins *I*-band, was chosen for observations at three epochs in 2004 (17 May, 2 August, 27 October) and with different telescope roll angles at each epoch (the +Y axis of the detector had PA 67.6˚, 93.5˚ and 222.6˚, respectively). The F814W data has a cumulative integration time of ~80 minutes. Additional follow-up data in the third epoch were obtained with the F606W filter ($\lambda_c = 591$ nm, $\Delta\lambda = 234$ nm), comparable to the Johnson-Cousins *V*-band, with +Y axis PA 222.6˚, 226.6˚, and 230.6˚ within the third epoch. For all epochs and wavelengths the main reference star for subtraction of the point-spread function (PSF) was HD 172167, which was also observed with a wide range of position angles, but always in telescope orbits immediately before or after the Fomalhaut observations. In the third epoch, a second PSF reference star (HD 210418) was also observed with the F606W filter. Short integrations produced a set of images where the unsaturated PSF core is visible at high signal to noise behind the occulting spot, giving the location of the stellar centroid to a fraction of a pixel (<25 mas). The stellar full-width at half-maximum is 63 mas (0.5 AU) in F606W and 72 mas (0.6 AU) in F814W.

Data reduction included the standard pipeline processing from the HST archive that produces bias-subtracted and flat-fielded image files. Bad pixels were replaced by interpolation using neighbour pixel values. Each Fomalhaut PSF was then subtracted iteratively using the reference star data. Registration was typically sensitive to 0.1 pixel relative shifts. The images were then corrected for distortion, resulting in images with 25 mas per pixel, registered to a common orientation, and median-combined. To test for nebulosity in our PSF reference star images, we performed the same data reduction steps, including registration to a common orientation, and further binned the data 16 x 16 pixels, with no detection of nebulosity in our PSF reference star fields. The Fomalhaut belt is detected at each epoch of observation, as are several background stars. To improve the signal-to-noise in the belt, we combined our F606W and F814W data using a weighted average [(0.69 x F606W + F814W) / 2]. For the combined data we use zero point equal to 24.4 mag. Assuming that Fomalhaut $m_V$ - $m_I$ = 0.09 mag, the stellar magnitude in the combined data is $m_{F606W+F814W}$ = 1.12 mag. Photometry on the belt was conducted using a circular aperture with diameter 1.25 arcsec centred on the brightest points along the belt. The images shown in Fig. 1 and 2 are not smoothed or binned.

The model belt shown in Fig. 1b and 3 has a volume number density distribution that varies as a function of radius and height above the midplane[13]. The vertical number density away from the midplane decreases exponentially, consistent with that derived for the edge-on disks β Pic and AU Mic[25,26]. The scale height flares as a power-law function of radius with exponent equal to 0.5. As described in the text, the model parameters (inner and outer radius, inclination to the line of sight, position angle, phase function asymmetry parameter, radial number density index, scale height, and flare index) were adjusted to provide a satisfactory subtraction of the ring in Fig. 1, and an approximate fit to the radial cut shown in Fig. 3. Since the belt may not have a uniform physical width around its circumference, the model belt chosen here is non-unique and may only apply to the section of the belt sampled in Fig. 2b. In Fig. 3, least squares exponential and power-law fits to the belt gradient in the inner edge (120-140 AU) are proportional to exp(0.08$r$) and $r^{10.9}$, respectively, where $r$ is in AU. The outer edge gradient (140-158 AU) is proportional to exp(-0.03$r$) and $r^{-4.6}$, respectively.

## Table 1. Properties of Trans-Planetary Belts

|  | Fomalhaut | Sun[1] |
|---|---|---|
| Age (Myr) | 200 ± 100[2] | 4600 |
| Spectral Type | A3V | G2V |
| Stellar mass (Solar mass) | 2.0 | 1.0 |
| Semi-major axis (AU) | 141 | 42, 48, 39[3] |
| Inner radius (AU) | 133 | 38, 25, 41[3] |
| Outer radius (AU) | 158 | 48, 53, 55[3] |
| Belt dust mass (gram)[4] | $1.1 \times 10^{26}$ | $6.0 \times 10^{22}$ |
| Albedo | 0.05 – 0.1 | 0.07 |
| Outermost planet (AU) | - | 30 |
| Symmetry Offset (AU) | 15.3 | 0.01[5] |
| Mean eccentricity | 0.11 | 0.09, 0.14, 0.36[3] |
| Mean Inclination (degrees) | 1.5 | 7.1, 9.8, 12.9[3] |

[1]Kuiper Belt properties from ref. 27, except for belt dust mass from ref. 23.
[2]Ref. 24;
[3]First, second and third entries are the mean values for Classical Kuiper Belt, the 2:1 resonance population, and the 3:2 resonance population, respectively.
[4]Ref. 10.
[5]Value for the Zodiacal dust cloud from ref. 28.